% [
                             % ]
\def\wdg{{\wedge}}                              % wedge product
\def\hepth#1{{\it hep-th/{#1}}}

\input harvmac
\input epsf

%%%%%%%%%%%%%%%%%%%%%%%%%%%%%%%%%%%%%%%%%%%%%%%%%%%%%%%%%%%%%%%%%%
\def\figin{\epsfcheck\figin}\def\figins{\epsfcheck\figins}
\def\epsfcheck{\ifx\epsfbox\UnDeFiNeD
\message{(NO epsf.tex, FIGURES WILL BE IGNORED)}
\gdef\figin##1{\vskip2in}\gdef\figins##1{\hskip.5in}% blank space
instead
\else\message{(FIGURES WILL BE INCLUDED)}%
\gdef\figin##1{##1}\gdef\figins##1{##1}\fi}
\def\DefWarn#1{}
\def\figinsert{\goodbreak\midinsert}
\def\ifig#1#2#3{\DefWarn#1\xdef#1{fig.~\the\figno}
\writedef{#1\leftbracket fig.\noexpand~\the\figno}%
\figinsert\figin{\centerline{#3}}\medskip\centerline{\vbox{\baselineskip12pt
\advance\hsize by -1truein\noindent\footnotefont{\bf Fig.~\the\figno:}
#2}}
\bigskip\endinsert\global\advance\figno by1}
%%%%%%%%%%%%%%%%%%%%%%%%%%%%%%%%%%%%%%%%%%%%%%%%%%%%%%%%%%%%%%%%%%

% \draftmode
\Title{ \vbox{\baselineskip12pt\hbox{hep-th/9708022, PUPT-1710 }}}
 {\vbox{
\centerline{ Scattering of zero branes off elementary  }  
\centerline{ strings in Matrix theory.} }}
%%% %
 \centerline{Rajesh Gopakumar and Sanjaye Ramgoolam}
 \smallskip
 \smallskip
 \centerline{Department of Physics, Jadwin Hall}
 \centerline{Princeton University}
 \centerline{Princeton, NJ 08544, USA}
 \centerline{\tt rgk,ramgoola@puhep1.princeton.edu}
%%% %%%
 \bigskip
 \bigskip
 \noindent

We consider the scattering of zero branes 
off an elementary string in Matrix theory or equivalently
gravitons off a longitudinally wrapped membrane.  
The leading supergravity result is recovered by 
a one-loop calculation in zero brane quantum mechanics. 
 Simple  
scaling arguments are used 
 to show that there are no further corrections 
at higher loops, to the leading term in the large impact parameter, 
 low velocity expansion. The  
mechanism for this agreement 
is identified in terms of properties of a recently discovered
boundary conformal field theory.

\Date{ August 1997}

\lref\cheptse{ I.Chepelev and A. Tseytlin, 
 {\it `` Long-distance interactions of D-brane bound states and longitudinal 5-brane in
     M(atrix) theory''}   \hepth{9701151} }

\lref\tseyt{A. Tseytlin, {\it ``Harmonic superpositions of M-branes,''}
 NPB475 (1996) 149, hep-th/9604035 } 

\lref\halyo{ E. Halyo,
 { \it ``M(atrix) black holes in five dimensions,''},   hep-th/9705167.  } 

\lref\shifva{V.A.Novikov, M.A.Shifman, A.I.Vainshtein, V.I. Zhakharov, 
 Fortschr. Phys. 32 (1984) 11, 585-622. }

\lref\Russo{J.G. Russo,
  {\it ``BPS Bound States, Supermembranes and T-Duality
  in M-Theory,''} \hepth{9703118} }

\lref\rustsey{ 
J.G. Russo, and A.A. Tseytlin, 
{\it `` Waves, boosted branes and BPS states in M-theory,'' 
\hepth{ 9611047} 
}}

\lref\mli{ M. Li and E. Martinec, {\it ``On the Entropy of 
Matrix Black Holes,''} \hepth{9704134} }
  
\lref\dvvb{ R. Dijkgraaf, H. Verlinde, E. Verlinde,
{\it ``5D Black Holes and Matrix Strings,''}
\hepth{9704018}
} 

\lref\DKPS{M.R. Douglas, D. Kabat, P. Pouliot and S. Shenker,
  {\it ``D-branes and Short Distances in String Theory,''}
     \hepth{9608024}}

\lref\gilmat{G. Lifschytz and  S. Mathur,
{\it `` Supersymmetry and Membrane Interactions in M(atrix) Theory,''} 
    \hepth{9612087} }

\lref\dps{ M. Douglas, J. Polchinski, A. Strominger,
{\it ``Probing Five-Dimensional Black Holes with D-branes,''}
\hepth{9703031}
 }
\lref\BFSS{T. Banks, W. Fischler, S.H. Shenker and L. Susskind,
  {\it ``M Theory As A Matrix Model: A Conjecture,''} \hepth{9610043}}

\lref\wati{W. Taylor,
  {\it ``D-brane field theory on compact spaces,''} \hepth{9611042}}

\lref\Gop{R. Gopakumar,
  {\it ``BPS States In Matrix Strings''},
  \hepth{9704030}}

\lref\BL{V. Balasubramanian and F. Larsen,
  {\it ``Relativistic Brane Scattering,''} \hepth{9703039}}

\lref\ggr{ O. Ganor, R. Gopakumar, S. Ramgoolam, 
{\it ``Higher loop effects in M(atrix) Orbifolds''}, 
\hepth{9706072}     } 

\lref\BSS{ T. Banks, N. Seiberg, S. Shenker,
{\it ``Branes from Matrices''},
\hepth{9612157} } 

\lref\bbpt{ K.Becker, M. Becker, J. Polchinski, A. Tseytlin, 
 {\it ``Higher Order Graviton Scattering in M(atrix) Theory'' }, 
\hepth{9706072}  }    

\lref\BB{K. Becker and M. Becker,
   {\it ``A Two-loop test of M(atrix)-theory''}, \hepth{9705091}}

\lref\Polpou{J.Polchinski, P.Pouliot,
   {\it ``Membrane Scattering with M-Momentum Transfer,''}
      \hepth{9704029}}

\lref\BFSS{T. Banks, W. Fischler, S.H. Shenker and L. Susskind,
  {\it ``M Theory As A Matrix Model: A Conjecture,''} \hepth{9610043}}

\lref\grt{ O. Ganor, S. Ramgoolam, W. Taylor,
{\it ``Branes, Fluxes and Duality in M(atrix)-Theory''},
\hepth{9611202}} 

\lref\Suss{L. Susskind,
   {\it ``Another Conjecture about M(atrix) Theory,''} \hepth{9704080}}

\lref\dvv{ R.Dijkgraaf, E. Verlinde, H. Verlinde,
{\it ``Matrix String Theory''}
\hepth{9703030}}
 
\lref\bbpt{ M. Becker, K. Becker, J. Polchinski, A. Tseytlin
{\it ``Higher Order Graviton Scattering in M(atrix) Theory''}
\hepth{9706072}} 

\lref\lenss{ L. Susskind as referenced in \bbpt }

\lref\ck{ C. Callan, I. Klebanov, {\it ``D-brane boundary state dynamics''}
\hepth{9511173}}

\lref\sv{ A. Strominger and C. Vafa
{\it ``Microscopic Origin of the Bekenstein-Hawking Entropy,''}
\hepth{9601029}} 

\lref\malda{ C. Callan and J. Maldacena,
{\it `` D-brane Approach to Black Hole Quantum Mechanics
``}\hepth{9602043}} 

\lref\limart{ M. Li and E. Martinec
{\it ``Matrix Black Holes''}
\hepth{9703211}}

\lref\doli{ M. Douglas and M. Li,
{\it `` D-Brane Realization of N=2 Super Yang-Mills Theory in Four
   Dimensions,}
\hepth{9604041}}

\lref\grgu{ M. Green and M. Gutperle 
{\it ``Light-cone supersymmetry and D-branes''}
\hepth{9604091}} 

\lref\hamo{ J. Harvey and G. Moore,
{\it `` Algebras, BPS States, and Strings
''} \hepth{9510182}}
 
\lref\baki{ C. Bachas and E. Kiritsis,
{\it ``  $F^4$ Terms in $N=4$  String Vacua''}
\hepth{9611205}}

\lref\sustalk{ L. Susskind, talk at Strings 97.  }
  
\lref\cheptseii{ I. Chepelev and A. Tseytlin, {\it ``Interactions of type IIB 
 D-branes from D-instanton matrix model,''} hep-th/9705120. }

 \newsec{ Introduction}

It is of great importance that we test the conjecture of M(atrix) 
theory \BFSS\ 
in different settings. This is specially true for the case with maximal 
supersymmetry
compactified on low dimensional tori where the
 conjecture is most unambiguous.
One might argue that the presence of
BPS states with the right masses etc. follow from supersymmetry. 
Similarly the agreement
between supergravity and Yang-Mills results for
 many of the scattering processes 
that have been studied is also a remnant
 property of the supersymmetry. 
In particular, the 
configurations have been mostly such that they reduce, in the limit
 when the radius
of the M-direction is small, to the annulus diagram in string theory
with simple D-brane boundary conditions at both ends.
 For this diagram
it is known  \DKPS , that the leading long distance, low velocity 
supergravity behaviour is reproduced by
the truncation to the lightest open string states. 

We would therefore like to have 
a number of cases where agreement between Matrix theory and
 supergravity does not rely
on this fact. One such instance is the calculation by Polchinski
 and Pouliot \Polpou . 
(Recently a calculation which tests a variant of the original 
conjecture \Suss\ has
also been performed \bbpt .) It is our intention in this paper 
to carry out a 
one loop calculation in the quantum mechanics which does 
{\it not} correspond to an 
annulus diagram in  a simple  D-brane string theory. 

The process that we are going to study is that of scattering gravitons (with
M-momentum ) off a longitudinal 
membrane in M-theory. In the string theory
 limit, the membrane becomes the fundamental
string while the 
graviton becomes a 0-brane. Thus our computation may be equivalently viewed
as a gauge theory calculation for 0-branes
scattering off the elementary string.
This involves a one-loop calculation about 
a non-trivial gauge field background.   
(Another example of such a  non-trivial 
  background was considered in \cheptse, and 
rather general techniques were developed in \cheptseii. ) 
Note also that our calculation is T-dual to the scattering
of a D-string probe off a system of D-string carrying momentum, and 
as such it is related to a certain limit of the scattering 
of D-string probe off black holes \dps. 

%%
%\ifig\disc{Zero brane string scattering}
%{\epsfxsize 3.0in\epsfbox{str.eps}}
%%

In the next section we consider the 
supergravity calculation at low velocities.
Section 3 deals with the one loop gauge 
theory calculation while Section 4 considers
some scaling arguments which strongly 
constrain the form of the one-loop 
answer and  show  that there are no corrections to our result.
Section 5 discusses the  mechanism for the agreement 
between 1-loop Yang Mills and supergravity  in terms 
of a boundary conformal field  theory. 
An appendix provides some details of the one loop calculation.  

%  We consider the scattering of zero branes 
%  off of elementary strings in Matrix theory. 
%  The supergravity result is recovered by 
%  a one-loop calculation in zero brane quantum mechanics. 
%  Unlike many examples studied so far, in this problem 
%  the string theory diagram which gives the leading 
%  order contribution to this process is not 
%  a tree level exchange of a closed string. 
%  Rather it is a process involving a worldsheet 
%  with one boundary and two closed string vertex operators. 
  
%%%%%%%%%%%%FIGURE%%%%%%%%%%%%%%%%%%%%%%%%%%%%%%%%%%

%\vskip2.5cm
%\epsfsize=4.0in \hskip 0cm \epsfbox{str.eps}
%\vskip0cm
%{\tenrm{
%FIGURE 1. 
% \hfil\break

%Leading diagram contributing to the 
%scattering of elementary string off 
%a zero brane. 

% }}

% \vskip2.0cm

%%%%%%%%%%%%%%%%%%%%%%%%%%%%%%%%%%%%%%%%%%%%%%%%%%

\newsec{The Phase Shift in Supergravity}

In this section we shall compute the leading contribution to the phase shift,
in the scattering of an eleven dimensional graviton off a longitudinally
wrapped membrane in supergravity.
Or equivalently, when the 
radius $R_{11}$ of the longitudinal M-direction becomes small,
that of a zero brane off an elementary string.
We also need our membrane 
to carry momentum in the M-direction, corresponding to the 
fact that in Matrix theory the object that 
we construct is built out of a large number
of zero branes. 
Though, as we will see,
this will not affect the leading velocity behaviour. 

We write down the
classical solution  of such a boosted membrane in 10+1D, 
which was found in \tseyt\ and reviewed for example in \Russo:
\eqn\memb{\eqalign{
ds^2 &= H^{-2/3} \lbrack -dt^2 + 
dy_1^2 + dy_{11}^2 + {Q_0 \over {r^6}}(dt - dy_{11})^2 
\rbrack  + H^{1/3} dx_i dx^i,\cr
C_3 &= H^{-1} dt\wdg dy_1\wdg dy_2,\cr
H &= 1 + {Q_2\over {r^6}}\cr
}}
where $Q_2= 8N_2 (2\pi)^2 (l_p)^6$ 
is the correctly normalised charge (see for example
\BL ),
$N_2$ is an integer and $i=2,3 \dots 9$

The geodesic equation for a graviton in this background is:
$$
|\vec{p}| {{dp_3}\over {dt}}
 = \Gamma_{300} |\vec{p}|^2 + 
\Gamma_{333} (p^3)^2 + \Gamma_{3,11,11} (p^{11})^2
+ 2 \Gamma_{3,0,11} |\vec{p}|^2
$$
with asymptotic momenta
$$
p_3 = {{N_0 v}\over {R_{11}}},\qquad
p_{11} = {{N_0}\over {R_{11}}},
\Longrightarrow
|\vec{p}| = {{N_0}\over {R_{11}}} (1 + v^2)^{1/2}.
$$
In the above, we have lowered all indices 
neglecting any metric dependence, justified
since we will only be  considering impact parameters which 
are large compared to Planck scale. 
Similarly we will only keep the leading velocity 
contributions.

Putting all this together gives the equation
\eqn\geod{\eqalign{
|\vec{p}| {{dp^3}\over {dt}}
&= - {{Q_2 x_3}\over {r^8}} \lbrack
3(p_3)^2(1-{Q_0 \over {r^6}}) - 2 |p_{11}|^2{Q_0 \over {r^6}}
\rbrack \cr
&-{{3Q_0x_3}\over {r^8}}\lbrack
(|\vec{p}|-p_{11})^2 +(p_{11})^2{Q_0 \over {r^6}}\rbrack
 = -{{3 Q_2 x_3}\over {r^8}} (p_3)^2,}}
where in the last line we have neglected terms 
which are subleading in powers of
${1\over r}$ as well as powers of velocity. 
We see that there is no $Q_0$ dependence
here at all. This can be understood since 
the terms proportional to $Q_0$
come from graviton-graviton scattering 
which exhibits a $v^4$ dependence. (The $r$   
dependence is like ${1\over {r^6}}$ rather 
than the familiar ${1\over {r^7}}$ because
these are zero branes bound to the  
membrane and hence have one homogeneous transverse 
dimension.) 

This corresponds to an effective force and potential:
$$
{{\partial V}\over {\partial x_3}} =
{{3 Q_2 x_3 v^2 N_0} \over {r^8 R_{11}}}
\Longrightarrow
V(r) = - {{Q_2 N_0 v^2}\over {2 r^6 R_{11}}}
$$
which corresponds to a  phase shift of 
\eqn\sugshf{ 
\delta = {{3\pi v Q_2 N_0}\over {16 R_{11} b^5}}=
    {  3 (2\pi)^3 v N_2N_0 l_p^6 \over {4b^5 R_{11} }}  
}

\newsec{ One loop calculation. } 

The matrix quantum mechanics contains, in the large N limit,
 a central charge in the
supersymmetry algebra corresponding to a longitudinally wrapped membrane \BSS. 
A BPS configuration carrying this charge may be easily constructed by realising
that this (winding) charge 
becomes momentum in a (T-dual) 1+1 dimensional Yang-Mills 
description. 
 
The $U(N)$ super-quantum mechanics Lagrangian is
\eqn\sqm{  L = {1\over{ 2R_{11} }} Tr \int dt \left\lbrack  (D_t X^i)^2 
+{  R_{11}^2\over {4\pi^2l_p^6}} [X_i,X_j]^2 + fermions \right\rbrack }
The Lagrangian for the 1+1 dimensional theory is:
\eqn\lagphys{ 
L = { 1\over 4\pi R_1 R_{11}  }Tr\int dt \int_{x_1=0}^{2\pi R_1}dx_1
\left\lbrack   (D_t X^i )^2 - ( D_{x_1} X^i )^2 
     + { R_{11}^2 \over { 4\pi^2l_p^6}} 
[X_{i}, X_{j} ]^2 + fermions  \right\rbrack.   }    

Here $i=2 \ldots 9$. This is obtained by the substitution   
$$
X_{1} = {1 \over { R_{11} } } \lbrack 
{ i \partial \over {\partial  x_1} } + A_1 \rbrack
$$
in the $U(N)$ super quantum mechanics Lagrangian, which implements
T-duality \BFSS\wati\grt. 
The variable $x_1$ is periodic  with 
period $2\pi R_1$.

It will be more convenient to work with rescaled variables where
\eqn\resmath{\eqalign{ 
 &  t_M = t_P { R_{11}\over {l_p^2}} (2\pi)^{-2/3} \cr 
  & X_M =  { (2\pi)^{-1/3}X_P \over {l_p}} \cr
  & (x_1)_M = (x_1)_P R_{11} \cr
  & (R_1)_M = ( R_1)_P R_{ 11} \cr 
 }}
   In these 
variables we have the Lagrangian: 
\eqn\lagphysi{ 
L = { 1\over 4\pi R_1   }Tr\int dt \int_{x_1=0}^{2\pi R_1}dx_1
\left\lbrack   (D_t X^i )^2 - ( D_{x_1} X^i )^2 +  
[X_{i}, X_{j} ]^2 + fermions  \right\rbrack,  }

There are BPS saturated states in the 
Yang-Mills which are essentially plane wave 
states, which in the  original quantum 
mechanical picture are longitudinal 
membranes wrapped on the T-dual circle $\tilde{X_{1}}$ \Gop. One such 
configuration with $n$ units 
of momentum along $X_{1}$ -- 
winding $n$ along $\tilde{X_{1}}$ is 
$$
X_2 = \sqrt {2  R_1 \over {  n  } }
               cos { n\over {R_1}} (x_1+ t).
$$
The normalization is such that this background
 has the right energy.

To calculate the leading order scattering 
of a zero brane off this background 
we evaluate the path integral at one loop around the background.  
\eqn\back{\eqalign{   
& X_2^{(1)} = A Cos { n\over {R_1}} (x_1+ t) \equiv 
               \sqrt {2  R_1 \over {  n  } }
               Cos { n\over {R_1}} (x_1+ t) \cr 
& X_3^{(2)}=  b \bf{I} \cr 
& X_4^{(2)}= v t \bf{I} \cr }}
Here we have decomposed the $X's$ in terms of 
$$X_{\mu} = X_{\mu}^{(1)} \oplus X_{\mu}^{(2)}$$ 
where the first operator acts on the space of functions
of the variable $x_1$ and the second operator is a $N_0 \times N_0 $ matrix 
where $N_0$ is the integer charge of the probe  zero brane. 
For simplicity we will work with 
$N_0=1$. 

As usual we expand the fields around the background keeping only the 
quadratic terms in the fluctuations. After taking into account
the ghosts and gauge-fixing terms \DKPS\gilmat, we 
separate the resulting determinants coming  from the
bosons, the fermions and the ghosts to the free energy, 
\eqn\sepcon{  W = W_B + W_G + W_F  }
In the notation of \cheptse
\eqn\cons{\eqalign{ 
&W_B = 
 Tr \log [ (b^2 + P^2)\delta_{\mu\nu}  + 2i F_{\mu\nu} ] \cr  
& W_G = -2  Tr \log (b^2 + P^2 )\cr 
& W_F = - \half  Tr \log (b^2 + P^2 + 
        {i\over 2} \Gamma_{ij}F_{ij} + \Gamma_4 v +
            \Gamma_1 F_{02}  ), \cr}}
where the non-zero elements of 
the matrices $F_{\mu\nu} (\mu,\nu=0\ldots
9)$ are
$$F_{04}=-F_{40}= v ; \ \ F_{12} = -F_{21} = \partial_{x_1}X_2 \ \
F_{02} = -F_{20} = \partial_{t}X_2.$$
Also
$$P^2 = \partial_t^2 -\partial_{x_1}^2 +
 v^2t^2 + b^2 + A^2 Cos^2 n(x_1+t). $$

$W_B$ (after cancelling out the terms in $W_G$) can be written out as  
\eqn\bos{\eqalign{ & W_B =  W_B^{(4)}   + W_B^{(i)} \cr 
&=  Tr \log [ (b^2 + P^2)\delta_{ab} + 
2i F_{ab} ] \cr 
& + 4 \sum_{n=0}^{\infty} 
Tr \lbrack\log ( b^2 + P^2 ) \rbrack,   \ \ (a,b = 0,1,2,4),  \cr }} 
where the  $W_B^{(4)}$ term comes from the kinetic 
terms for $A_0, X_1, X_2, X_4$, and 
$W^{(i)}$ is related to  
the remaining six bosonic fields less two ghost
degrees of freedom.

It is convenient to take a derivative: 
\eqn\expa{
{dW_B \over {db^2} }=  { dW^{(4)}\over {db^2}} + { dW^{(i)}\over {db^2}}
 }
We will outline the calculation of the first term 
on the RHS above. The calculation of $W^{(i)}$ and $W^{(F)}$ 
is very similar. 
\eqn\calci{\eqalign{ 
 &  { dW^{(4)}\over {db^2}} = Tr ( {1\over {H_0+H_1}} ) \cr 
 & = Tr  { 1\over H_0} - Tr (  { 1\over H_0}H_1
 { 1\over H_0})  +  
Tr ( { 1\over H_0}H_1{ 1\over H_0}H_1{ 1\over H_0}) 
 + \cdots \cr}}
The trace includes integrals over $x_1,t$ space as well as 
a sum over a discrete index running through $0,1,2,4$. 
We have separated the kinetic operator into 
a piece $H_0$ which can be diagonalized exactly, and a piece 
$H_1$ which we treat pertubatively. 
\eqn\sep{ \eqalign{ 
&  H_0 = (b^2 + \partial_{t}^2 + 
 v^2t^2 - \partial_{x_1}^2 + 2iF_{04}e_{04}  ) \cr 
& H_1 = A^2 Cos^2 p(x_1 + t) + 2i F_{12}e_{12} + 2iF_{02}e_{02} \cr}}
The matrix $ e_{ij}$ has non-zero entries 1 in the i'th row and $j$'th column
and $-1$ in $j$'th row and $i$'th column. 
The dots in \calci\ represent terms which are down 
in the large $b$ expansion. 
Evaluation of these terms requires 
consideration of commutators, for example : 
\eqn\com{ \eqalign{ 
& Tr (  { 1\over H_0}H_1 { 1\over H_0 } ) = Tr  (  { 1\over H_0^2}H_1 )
 + Tr  ( { 1\over H_0^3} [H_0,H_1] )  + 
Tr  ( { 1\over H_0^4} [H_0,[H_0,H_1]] ) + \cdots }}
The last term and higher are 
of form  $1\over b^7( 1 + { v\over b^2} + \cdots )$
and cannot contribute to the terms of interest. 
The second term gives vanishing contribution when evaluated. 

In case of $W^{(F)}$ we again treat the $A^2Cos^2$ term as well 
$F_{12}$ and $F_{02}$ as perturbation. Now the calculation involves
an extra trace over spinor indices. 

The final answer,  is 
$$ W= { 3\pi  v n  \over {2b^5}}.$$ 
After converting to the physical variables, using \resmath, we get,   
\eqn\fnal{ 
 W =  { 3 vn (2\pi)^3 l_p^6 \over {4b^5 R_{11}}, }      } 
in agreement with \sugshf\  for $N_2=n$ and $N_0=1$. 
The $N_0$ dependence also agrees as discussed 
at the end of the next section.    
Some more details about the calculation are given in the appendix.

\newsec{ General properties of  the loop expansion}
 
Some elementary scaling arguments
can be used to derive properties of the 
loop expansion around backgrounds of interest in Matrix theory. 
These arguments have been developed in the simpler cases 
of flat space and non-compact orbifolds in 
\ggr\bbpt\lenss.
We write down the physical Lagrangian, and then change variables 
to a form where the loop expansion parameter is manifest.
The physical Lagrangian is: 
\eqn\lagphys{ 
L = { 1\over 4\pi R_1 R_{11}  }\int dt  \int_{x_1=0}^{2\pi R_1} 
dx_1 \lbrack (D_t X^i )^2 - ( D_{x_1} X^i )^2 
     + { R_{11}^2} [ X_{i}, X_{j} ]^2 + i \Theta^TD_t \Theta 
     - \Theta^T\Gamma^i[X_i, \Theta] \rbrack  } 
We are using units where  the eleven dimensional 
Planck length is set to $1$. 
This is obtained from the super quantum mechanics Lagrangian
by the substitution
\eqn\relzer{
 X_{1} = { 1 \over R_{11}}
(  i {\partial \over {\partial x_1}} + A_1). }

We can rewrite this in terms of variables $t_h, (x_1)_h, X_h$
which are related to the above variables by: 
\eqn\chngva{\eqalign{  
&   t_h = {t\over  R_1} \cr
&   (x_1)_h = {x_1\over { R_1}} \cr 
&   X_h^{i} = X^{i}R_1R_{11} \cr
& \Theta_h = (R_1R_{11})^{ 3/2 }  \Theta \cr  }}
In these variables the Lagrangian takes the form (numerical coefficients 
 will not be important in this section) : 
\eqn\laglup{
 { 1\over {2R_1^3R_{11}^3}  } \int dt   \int_{x_1=0}^{2\pi} d(x_1)_h
   \lbrack (D_{t_h} X_h^i )^2 - 
( { \partial X^i\over {\partial (x_1)_h}} )^2 
 + R_{11}^2 [ X_{i}, X_{j} ]^2 +\Theta_h^TD_t \Theta_h 
      -  \Theta_h^T \Gamma_I[X_h^I, \Theta_h ]  
\rbrack }

The backgrounds that we consider are therefore related
by the equations: 
\eqn\backgd{\eqalign{ 
& b_h = b R_1 R_{11} \cr 
& v_h = v R_1^2 R_{11}\cr }}
In these h-variables,  where the loop expansion is manifest,   
the $X_2$ background in \back\
 is proportional to $(R_1R_{11})^{3/2}$. The L-loop result
has the form $ ( R_1^3R_{11}^3)^{L-1}  f_L ( v_h, b_h, R_1R_{11}) $, 
 where $f_L$ contains positive integral powers 
 of $(R_1R_{11})^3$.  
So not all the variables $b,v,R_{11}, R_1$ contribute 
independently at a fixed loop level. In these variables 
the L-loop result takes the form 
\eqn\llup{ 
\delta_{L} (b, v, R_{11}, R_1 ) =   ( R_1^3R_{11}^3)^{L-1} 
g_L ( b_h, v_h , R_1R_{11} ) }
At one loop this allows terms of the 
form 
\eqn\alform{\eqalign{ 
    { v \over {b^5 R_{11} }} &= 
{ v_h\over {b_h}^5} (R_1R_{11})^3 ,    \cr 
    { v^3R_1 \over {b^5 R_{11}^2} } &=  { v_h^3 
\over { b_h^5 }} 
\cr }}
The first term has precisely the $R_{11}$, $R_{1}$ dependence 
of the supergravity phase shift for zero brane off elementary string. 
 The second term has the right form for zero branes off zero branes 
smeared along the one dimension of the elementary string. 

We can obtain further insight into the form of 
the phase shift computed from Yang Mills, if we 
organize it in the number of loops as well as the number 
of insertions $2K$ of the $X_2$ background. From
\back, \backgd, and \chngva\ it follows that 
$X_2^h \sim (R_1R_{11})^{3/2} {1\over \sqrt{n}} \cos n((x_1)_h+t_h)$.
It is significant that the only the combination 
$R_1R_{11}$ appears and sits   
outside the cosine. 
Therefore loop diagrams with $L$ loops and $2K$ insertions 
of the $X_2$ background should have the form 
$(R_1R_{11})^{3(L-1+K)}f_{L,K} (v_h, b_h) $ in the $h$-variables
\eqn\formexp{ \delta_{L,2K} (v,b,R_1,R_{11}) =  
 (R_1R_{11})^{3(L-1 + K)} f( v_h , b_h  ) }
(We expect that the 
terms coming from $2K$ insertions 
go like $n^{K}$.)
It follows from  \formexp\ that 
the  ${v\over {R_{11}b^5}}$ cannot get any 
corrections from generic  higher loops or higher insertions. 
The only term that could perhaps correct it is 
the $L=2$, $K=0$ term but this would have no dependence on 
the elementary string charge, and so would be due to 
scattering of zero brane probe off the zero branes 
bound to the elementary string. But the order $v$ term 
in zero-zero scattering should vanish by supersymmetry 
and this has been explicitly demonstrated in 
\BB.   Therefore the $v\over b^5$ term we are comparing 
with supergravity can only come in Yang Mills from a one-loop 
calculation with an insertion of two powers of $X_2$.

Finally we comment on the range of parameters where the 
expansion is valid. 
Even at one loop there can be corrections with 
higher powers of 
$R_1R_{11}$, ${v_h \over{ b_h^2}} 
=  {v\over {b^2R_{11}}}$ and $b_h = bR_1R_{11}$. 
The expansion is valid in the 
regime where: 
\eqn\expva{\eqalign{ & {1\over  {R_1R_{11}}}
                              \sim    R_1^{(0)} \gg 1 \cr 
                      &  { b\over {R_1^{(0)}} } \gg 1 \cr 
                      &  {v\over b^2R_{11} }  \ll 1 \cr
}} 
Here ${R_1^{(0)}} = {1\over { R_1R_{11}}} $ is the 
physical size of the circle on which the 
elementary string is wrapped 
in M theory. 

When we allow  the $X^{(1)}$ block to act on 
an N-dimensional space tensored with functions of $x_1$, 
 and the second block to act on a $N_0$ dimensional space, 
 the above discussion is modified as follows.
The $N= N_0 = 1$ system considered so far corresponds
to having one zero brane on a circle bound to the elementary string
interacting with one zero brane probe. In general 
we have $N$ zero branes bound to the elementary string, and 
the probe can have charge $N_0$. 
The normalization of the perturbation gets modified by a factor
$\sqrt{ 1\over N}$. The one loop calculation gets a factor 
$NN_0$ from the multiplicity of off-diagonal fields. So the
leading answer is modified by a factor proportional to $N_0$, 
as expected from supergravity. 
 In general the 1-loop answers with $2K$ insertions 
pick up a factor of $N^{{ K}}$. 
At higher loops the leading large $N$ 
behaviour is $N^{L-1 + K}$.

\newsec{ String Diagrams and the Yang Mills calculation} 
  Let us review a few features of the 
 simpler one-loop calculations which have been 
 used to test of Matrix Theory. The simplest 
 case of scattering of zero branes off zero branes
 relies on the agreement between the short and long distance 
 behaviours of   closed string exchange amplitudes 
 (cylinder diagram) \DKPS. We would like to ask what kind 
 of property of string amplitudes is responsible 
 for the agreement between the 1-loop Yang Mills 
 result and the scattering of zero branes off 
 an elementary string source found in supergravity.   
 Conversely, we can obtain information about  the string 
 theory formulation of the problem of scattering zero branes 
 off an elementary string background, by analysing the 
 Yang Mills Matrix theory calculation.

 The system of   zero-brane with elementary string is 
T-dual to D-string carrying momentum. 
This system can be described by 
adding to  the  flat space closed string worldsheet action 
 boundary operator \ck\ of
the form: 
 \eqn\pert{ 
|B> = e^{ {\cal O} } |B_0> }
Here $|B_0>$ is the boundary operator 
for a D-string alone. The exponential 
is responsible for turning on momentum. 
We can expand the exponential and think of 
$B$ as creating a  boundary with D-string boundary conditions
 and the powers of $\cal O$ as insertions at the boundary. 
In our case the relevant form of ${ \cal O}$ has been considered 
by Callan and Klebanov in \ck, who showed 
that it defines a valid conformal field 
theory: 
\eqn\optor{ { \cal O}  = \int d\sigma 
a_p e^{ipX^{+}} ( \epsilon^j  {  \partial \over \partial {\tau}} X^j + 
 i p \psi^+\psi^j )  + c.c \rbrack  
 }

In the Yang Mills calculation, the diagram that leads to the 
leading effect is of the form shown in Fig.1. 
\ifig\disc{The one loop diagram with an insertion}
{\epsfxsize 3.0in\epsfbox{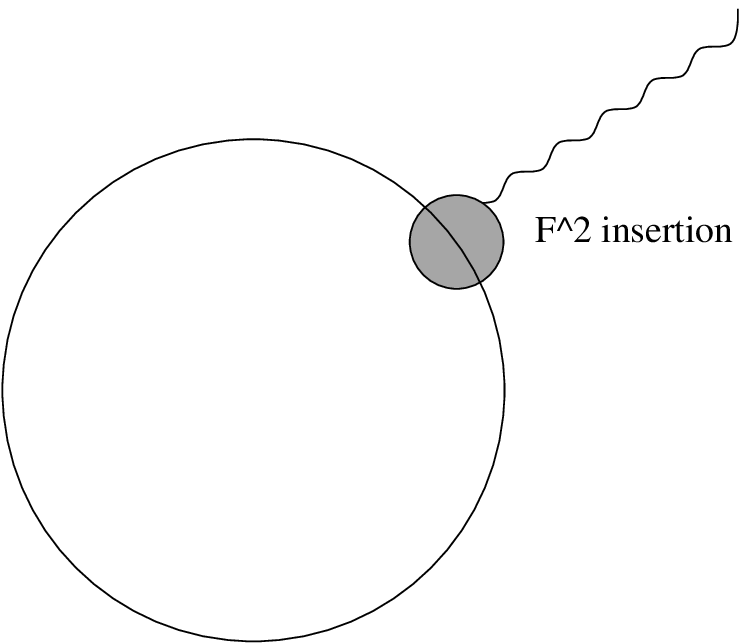}}
We have drawn a composite operator $F^2$
because the contributing terms involves
the perturbation $\bar X_2$ in the form 
$ \int dx ( \partial_t \bar X_2 (x) )^2$.
This clearly corresponds to a degeneration of the 
string theory diagram with two insertions 
of the ${ \cal O}$ operator at the boundary, 
as shown in fig. 2. 
 \ifig\disc{Degeneration of the zero brane string scattering diagram}
{\epsfxsize 3.0in\epsfbox{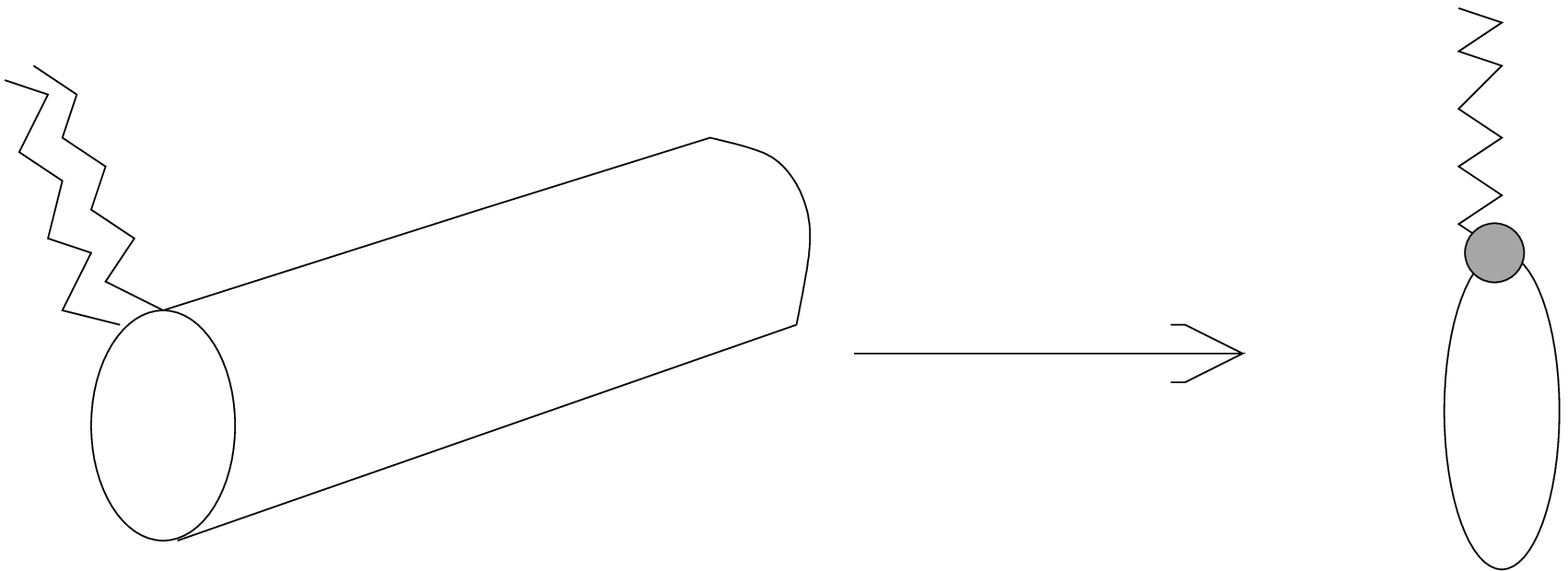}}

So the mechanism for the agreement between 
Matrix theory and supergravity 
is again the fact that for this cylinder 
diagram the contributions of the 
long multiplets cancel, so that it suffices to look at the 
lightest open string modes 
\doli\grgu\hamo\baki. 
It will be very interesting 
to understand in some generality what class of boundary CFTs
allows such a decoupling. In particular it will be interesting to see
if they include those  describing 
systems of D-branes, which are relevant in black hole physics
\sv\malda.
 This should shed further light on the
Matrix theory approaches to black holes
that have been started in \dps\limart\dvvb\mli\halyo. 

To describe just the elementary string  
rather than elementary string with zero branes, 
(or in the T -dual version plane wave without D-string) 
 a guess is that we just replace $|B_0>$ in the boundary operator
\optor\ by the ordinary vacuum of the bulk CFT, rather than 
the D-string boundary state.

\newsec{Discussion and conclusion}
We have studied scattering off elementary strings 
bound to zero branes in Matrix theory. 
We did  a one loop calculation in the 
superquantum mechanics to get the leading 
long distance, low velocity term in the phase shift. 
 We argued, using scaling and 
supersymmetry, that the loop expansion does not correct 
this leading phase shift.   

A point to be noted is that in the supergravity calculation
we used a simple solution which is expected \rustsey\ to have 
 the correct long distance properties of a longitudinal 
membrane carrying momentum.   
 A choice of profile may be necessary for precise
 matching at higher orders in the long distance expansion.

Systems with momentum, D1-brane and D5-brane  charge 
are of direct interest in the Matrix theory
description of black holes. 
We have seen in this paper that backgrounds  with 
somewhat non-trivial time dependence are 
correctly incorporated  in Matrix theory. 
As in \ggr\bbpt\lenss\  we have seen here that 
simple dimensional 
arguments give a lot of information about 
the phase shifts. We expect this to be quite 
general. Further it will be interesting 
to explore  the detailed connection 
between Matrix Superquantum mechanics and degenerations 
 of diagrams in open-closed string theories.

Several other interesting questions
can be addressed in these models. It 
is important to know how far we can go 
in recovering the long distance expansion 
of scattering amplitudes without 
knowing the detailed form of the 
zero brane bound state wavefunctions. This 
has been discussed to some extent  in \ggr\bbpt.
It is also important to understand possible 
relations of  the Matrix loop expansion to the 
non-compact limit of M-theory \sustalk.   

%=======================================================================%
% Acknowledgments
%=======================================================================%
 \bigbreak\bigskip\bigskip
\centerline{\bf Acknowledgments}\nobreak
It is a pleasure to thank V. Balasubramanian,
 O. Ganor, Z. Guralnik, A. Hashimoto,  F. Larsen,
V. Periwal, L. Susskind,  O. Tafjord and  A.Tseytlin
 for helpful conversations. This  work was supported 
by NSF Grant PHY96-00258 and DOE Grant DE-FGO2-91-ER40671. 

{\bf Appendix}

Here we put together some details of the one loop computation. 
The explicit evaluation of the determinants is done
in Euclidean space by the substitution $t \rightarrow it$, 
$v\rightarrow -iv$. 
For instance, in calculating $W^{(4)}$ using \calci\ and \sep\ we  
had to deal with $Tr  { 1\over H_0}$, $Tr  (  { 1\over H_0^2}H_1 )$,  
$Tr  ( { 1\over H_0^3} [H_0,H_1] )$ and $Tr  (  { 1\over H_0^3}H_1^2 )$. 

Thus, diagonalising the $4\times 4$ matrix in $H_0$, gives 
\eqn\ho{\eqalign{Tr  { 1\over H_0} =& 2 tr {1\over b^2 + \partial_{t}^2 
+v^2t^2 - \partial_{x_1}^2} \cr +& tr {1\over b^2 + \partial_{t}^2 
+v^2t^2 - \partial_{x_1}^2 + 2v} \cr  +& tr {1\over b^2 + \partial_{t}^2 
+v^2t^2 - \partial_{x_1}^2 - 2v}  }}
where the trace $tr$ is only over $x_1,t$ space. 
Each of these traces can be evaluated and written in the form 
\eqn\hotrace{ \sum_{n=0}^{\infty} \int dp {1 \over p^2 + b^2 +iv(2n+a)}}
with $a=\pm 1,3$.

The integral and sum can then be performed by using 
\eqn\trasum{\eqalign{ \sum_{n=0}^{\infty}{1 \over (b^2 + iv(2n+a))^{1/2}}
=&{1\over \pi^{1/2}b}\sum_{n=0}^{\infty}\int_{-\infty}^{\infty}dx
e^{-[1+(2n+a)\gamma]x^2}\cr =& {1\over \pi^{1/2}b}\int_{-\infty}^{\infty}dx
e^{-x^2}{e^{-(a-1)\gamma x^2}\over 2sinh(\gamma x^2)}.}}
Here $\gamma = {iv \over b^2}$.

Similarly, in evaluating $Tr  (  { 1\over H_0^2}H_1 )$ 
one first diagonalises the
$4\times 4$ matrix in $H_0$ and conjugates $H_1$ by the appropriate unitary
matrix. One obtains
\eqn\hoho{\eqalign{Tr  (  { 1\over H_0^2}H_1 )=& 2 
tr {1\over (b^2 + \partial_{t}^2 
+v^2t^2 - \partial_{x_1}^2)^2}A^2 Cos^2 {n\over {R_1}}(x_1+t) \cr
+&  tr {1\over (b^2 + \partial_{t}^2 
+v^2t^2 - \partial_{x_1}^2 +2v)^2}A^2 Cos^2 {n\over {R_1}}(x_1+t) \cr
+&  tr {1\over (b^2 + \partial_{t}^2 
+v^2t^2 - \partial_{x_1}^2 -2v)^2}A^2 Cos^2 {n\over {R_1}}(x_1+t).}}

The trace is carried out similarly. There is an average over the
 $\cos^2$ term as
well. There are sums of the 
form $\sum_{n=0}^{\infty}
{1 \over (b^2 + iv(2n+a))^{3/2}}$, 
which are evaluated by differentiating \trasum\
 with respect to $b^2$.
The term $Tr  ( { 1\over H_0^3} [H_0,H_1] )$ vanishes 
because $( \partial_{x_1}^2 - \partial_t^2) A^2 \cos^2 p(x_1+t)$ 
vanishes and because $\partial A^2 \cos^2 p(x_1+t)$ gives zero 
when integrated over $x$. 

The crucial contribution comes from the term $Tr  (  { 1\over H_0^3}H_1^2 )$.
Diagonalising as before gives us 
\eqn\hohoho{\eqalign{Tr  (  { 1\over H_0^3}H_1^2 ) =& 
tr {1\over (b^2 + \partial_{t}^2 +
v^2t^2 - \partial_{x_1}^2)^3}(2A^4Cos^4 {n\over {R_1}}(x_1-t)-4
(F_{02}^2+2F_{12}^2)) \cr +& tr {1\over (b^2 + \partial_{t}^2 
+v^2t^2 - \partial_{x_1}^2 +2v)^3}
(A^4Cos^4 {n\over {R_1}}(x_1-t)-2F_{02}^2) \cr
+& tr {1\over (b^2 + \partial_{t}^2 
+v^2t^2 - \partial_{x_1}^2 -2v)^3}
(A^4Cos^4 {n\over {R_1}}(x_1-t)-2F_{02}^2)}}

Finally adding these up gives 
\eqn\deriv{\eqalign{{dW^{(4)}\over d b^2} =& {1 \over 2\pi^{1/2}b}
\int_{-\infty}^{\infty}dx
e^{-x^2}{[1+cosh(2\gamma x^2)]\over 2sinh(\gamma x^2)} \cr 
+& {A^2\over 4\pi^{1/2} b^3}\int_{-\infty}^{\infty}dx x^2 e^{-x^2}
{[1+cosh(2\gamma x^2)] \over sinh(\gamma x^2)} \cr
+& { 3A^4 \over 32\sqrt{\pi} b^5} 
\int dx { x^4 e^{-x^2 } \over {sinh ( \gamma x^2 )}} 
\lbrack 1 + Cosh ( 2\gamma x^2) \rbrack \cr 
-&{\langle F_{12}^2 \rangle\over 2\pi^{1/2}b^5 }
\int_{-\infty}^{\infty}dx x^4 
e^{-x^2}
{[1-cosh(2\gamma x^2)] \over sinh(\gamma x^2)} \cr}}
where $\langle F_{12}^2 \rangle = \int_{0}^{2\pi R_1}dx_1 F_{12}^2 $. 

The calculation for $W^{(i)}$ 
can be performed in a similar manner. We obtain
\eqn\wi{\eqalign{{dW^{(i)}\over db^2}=& 
{1 \over \pi^{1/2}b}\int_{-\infty}^{\infty}dx
{e^{-x^2}\over sinh(\gamma x^2)} \cr
+&{A^2\over 2\pi^{1/2} b^3}
\int_{-\infty}^{\infty}dx x^2 {e^{-x^2}\over sinh(\gamma x^2)}\cr
+&{3A^4\over 16\pi^{1/2}b^5} 
\int_{-\infty}^{\infty}dx x^4{e^{-x^2}\over sinh(\gamma x^2)}.}}

Then there is the fermion  
contribution which is evaluated in a way very similar to that 
of $W^{(4)}$. Now the trace includes 
one over the $16\times 16$ $\Gamma$ matrices. Most of the 
terms vanish since the trace 
of a product of one, two and three distinct $\Gamma$ matrices
is zero. This gives as a result 
\eqn\wf{\eqalign{ {dW_F\over db^2}=& 
-{2 \over \pi^{1/2}b}\int_{-\infty}^{\infty}dx
{e^{-x^2}\over sinh(\gamma x^2)} cosh(\gamma x^2) \cr
-&{A^2\over \pi^{1/2} b^3} \int dx x^2 
{e^{-x^2}\over sinh(\gamma x^2)}cosh(\gamma x^2)\cr
-&{3A^4\over 8 \pi^{1/2} b^5}
\int_{-\infty}^{\infty}dx x^4{e^{-x^2}\over sinh(\gamma x^2)}
cosh(\gamma x^2)\cr
+&2{\langle (F_{12}^2-F_{02}^2)\rangle
\over \pi^{1/2} b^5}
\int_{-\infty}^{\infty}dx x^4{e^{-x^2}\over sinh(\gamma x^2)}
cosh(\gamma x^2).}}

In the  sum ${dW^{(4)}\over db^2}+ {dW^{(i)}\over db^2}+
 {dW^{(F)}\over db^2}$,  the only non-vanishing contribution 
to order ${v\over b^7}$ is  seen to come from the 
$F_{12}^2$ term  in \deriv.
\eqn\dw{ {dW \over db^2} =  { 15\pi v \over 4b^7  }} 
 The rest of the terms 
start at order $v^3\over b^7$. 

\listrefs
\end